\documentclass{aa}

\usepackage{times}
\usepackage{graphicx}
\usepackage{xspace}
\usepackage{epsfig}
\usepackage{natbib}
\usepackage{rotating}

\newcommand{\lgLrat}{\ensuremath{\log{(L_{\rm x}/L_{\rm bol})}}\xspace}

\begin{document}

\title{X-ray detection of the substellar twin 2MASS\,J11011926-7732383\,AB} 

\author{B. Stelzer\inst {1} \and G. Micela\inst {1}}

\offprints{B. Stelzer}

\institute{INAF - Osservatorio Astronomico di Palermo,
  Piazza del Parlamento 1,
  I-90134 Palermo, Italy \\ \email{B. Stelzer, stelzer@astropa.unipa.it}} 

\titlerunning{X-ray detection of the substellar twin 2M\,1101-7732}

\date{Received $<$28-03-2007$>$ / Accepted $<$05-06-2007$>$}

\abstract
{2MASS\,J11011926-7732383\,AB (hereafter 2M\,1101\,AB), 
located in the Cha\,I star forming region, is a rare wide-separation brown dwarf binary. 
Being spatially resolvable in many wavebands, 
it is a unique target for studying the properties of substellar twins. 
}
{
Here, we exploit the coeval pair 2M\,1101\,AB to examine 
the influence of physical parameters (mass, bolometric luminosity and effective temperature) 
on X-ray emission from substellar objects. 
}
{
We determine the X-ray properties of 2M\,1101\,A and~B using XMM-Newton and Chandra observations. 
}
{
The spatial resolution of XMM-Newton is not 
sufficient to separate contributions from the two components in the binary. The X-ray source
detected with XMM-Newton has a column density compatible with the infrared extinction of 
component A. 
On the other hand, 
the binary is resolved with Chandra, and the bulk of the X-ray emission is clearly associated 
with the photospherically cooler component B. 
These apparently contradictory results point at strong variability of 2\,M1101's
X-ray emission. Combined with previous sensitive X-ray observations
from low-mass members of Cha\,I, we find a decline of X-ray luminosity 
with decreasing (sub)stellar mass that is typical for star forming regions.  
}
{
2M\,1101\,B is the coolest (spectral type M8.25) 
and least massive brown dwarf of Cha\,I detected in X-rays so far. 
It is also among the youngest ($\sim 1$\,Myr) substellar Cha\,I members, and therefore
relatively luminous.  
Most bona fide brown dwarfs of Cha\,I have remained below the sensitivity limits of available
X-ray observations, because of their low luminosity associated with higher age.  
}


\maketitle

\section{Introduction}\label{sect:intro}

The binary brown dwarf 2MASS\,J11011926-7732383\,AB (hereafter 2M\,1101) 
was discovered by \cite{Luhman04.3} during a search for substellar objects in the
Cha\,I star forming region. 
Upon its discovery it was the first known {\em wide} brown dwarf binary,
and as such it constitutes an important discriminant for the formation mechanisms of brown dwarfs: 
the wide binary separation of $1.44^{\prime\prime}$ 
\citep[corresponding to $240$\,AU at the distance of $168$\,pc;][]{Whittet97.1, Wichmann98.1}
practically excludes the ejection scenario that predicts a maximum separation of $\sim 10$\,AU for
substellar binaries expelled from a multiple system without being disrupted \citep{Bate02.1}. 

\begin{figure}[h]
\begin{center}
\resizebox{8.5cm}{!}{\includegraphics{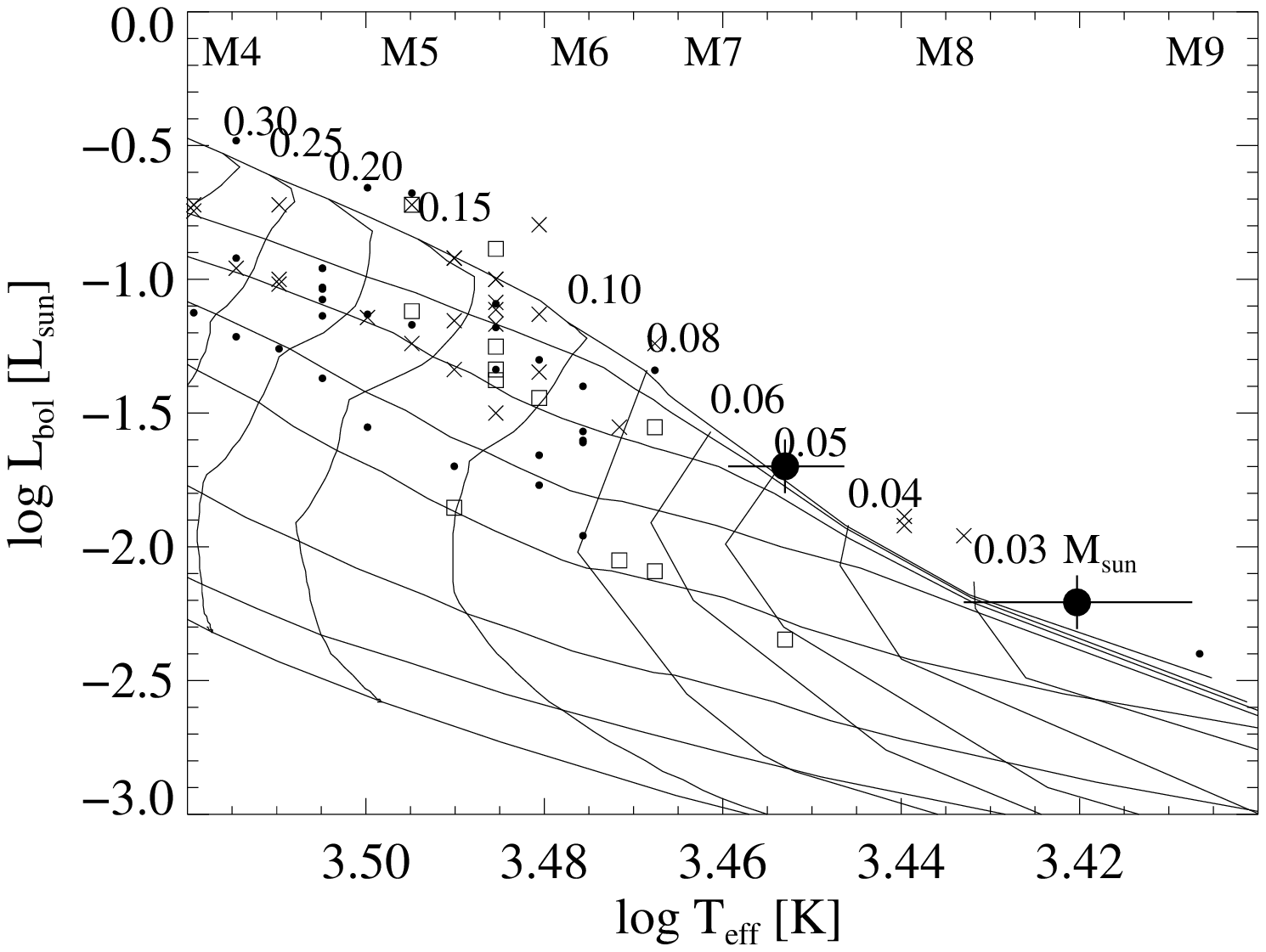}}
\caption{HR diagram for low-mass members of Cha\,I; data from \cite{Luhman04.1} and \cite{Comeron04.1};
models from \cite{Baraffe98.1} for
$M \geq 0.1\,M_\odot$ and from \cite{Chabrier00.2} for $M < 0.1\,M_\odot$. The tracks
are labeled by their mass, and the substellar limit corresponds to $0.07\,M_{\rm \odot}$. 
Isochrones represent from top to bottom $\log{{\rm Age}\,{\rm [yrs]}} = 6.0, 6.3, 6.5, 6.7, 7.0, 7.5, 8.0$ and $9.0$. 
Plotting symbols:
{\em crosses} - Cha\,I members detected in X-rays \citep{Stelzer04.2, Feigelson04.2, Robrade07.1};
{\em squares} - Cha\,I members not detected in X-rays; 
{\em small circles} - Cha\,I members not observed in X-rays. 
{\em large filled circles} - the brown dwarf binary 2M\,1101\,A and 2M\,1101\,B 
\protect\cite[stellar parameters from][]{Luhman04.3}, 
whose X-ray data are discussed in this paper.
The least massive brown dwarf in Cha\,I, OTS\,44 \citep[see][]{Luhman04.4}, is off the plotted
range to the right, and not detected in X-rays.}
\label{fig:hrd}
\end{center}
\end{figure}

According to \cite{Luhman04.3}, the two components in the 2M1101 system have spectral types M7.25 and 
M8.25, and bolometric luminosities of $0.02$ and $0.0062\,L_\odot$, respectively. 
Fig.~\ref{fig:hrd} shows the position of 2M\,1101 in the HR diagram, together with 
other low-mass Cha\,I members from the catalogs of
\cite{Luhman04.1} and \cite{Comeron04.1}. The 2M1101 binary (large filled circles in Fig.~1)
is coeval with an age of $\sim 1$\,Myr on the evolutionary tracks by \cite{Chabrier00.2}.
The probability for the two being a chance projection has been
estimated by \cite{Luhman04.3} to be extremely low ($5 \cdot 10^{-5}$). 
The individual masses of the two components according to the Chabrier et al. models are
$0.05\,M_\odot$ and $0.025\,M_\odot$ for 2M\,1101\,A and~B, respectively, well below
the transition to the substellar regime. The physical parameters of 2M\,1101 are summarized
in Table~\ref{tab:optir}. 

\begin{table}
\begin{center}
\caption{Physical parameters of the 2M\,1101 system relevant to this study; data from \cite{Luhman04.3}.}
\label{tab:optir}
\begin{tabular}{llrrrr}
\hline
Component & SpT    & $T_{\rm eff}$ & $L_{\rm bol}$ & $A_{\rm J}$  & $A_{\rm V}^{(1)}$ \\
          &        & [K]           & [$L_\odot$]   & [mag]        & [mag]             \\
\hline
A         & M7.25  & $2838$        & $0.020$       & $0.45$       & $1.6$         \\
B         & M8.25  & $2632$        & $0.0062$      & $0$ ($<0.2)$ & $0$ ($<0.7$)  \\
\hline
\multicolumn{6}{l}{$^{(1)}$ from $A_{\rm J}$ using the extinction law of \protect\cite{Rieke85.1}.}\\
\end{tabular}
\end{center}
\end{table}

Evidently, 2M\,1101 is a unique target for studying the properties of substellar twins, i.e.
coeval brown dwarfs of slightly different effective temperature, luminosity, and mass. 
Its wide separation allows to resolve the two components with present-day instrumentation
in many wavelength regimes, including the soft X-ray band.  

X-ray emission from 
late-type stars 
is a well-known signature of magnetic activity 
\citep{Rosner85.1}. 
Young pre-main sequence (pre-MS) stars show much higher levels of X-ray luminosity than
evolved stars in open clusters and in the field \citep[e.g.][]{Feigelson99.1}. 
The empirical saturation level for the X-ray luminosity of active stars is \lgLrat $~\approx -3$,
but  for given $L_{\rm bol}$ the spread reaches about $2$\,dex below this threshold.
According to recent observations the $L_{\rm x} - L_{\rm bol}$ relation seems to hold
even in the substellar mass regime \citep{Preibisch05.2, Grosso06.1}. 

The universality of the $L_{\rm x} - L_{\rm bol}$ relation can not be taken for granted, 
especially in view of the fact that pre-MS stars as well as brown dwarfs are fully convective and the 
standard solar-like $\alpha\Omega$-dynamo, 
which requires a transition between a radiative and a convective layer, is not expected to work.
Alternative field generating mechanisms involving turbulence 
have been invoked to explain X-ray emission from the fully convective lowest mass
stars at the bottom of the MS and from brown dwarfs \citep{Durney93.1}. 
The effects of a change in the dynamo mechanism
are both difficult to predict and observationally poorly constrained.
While there is consensus that the X-ray luminosity declines towards the substellar boundary,
it is unclear which role different physical parameters besides $L_{\rm bol}$ (e.g. rotation,
effective temperature, mass, age, binarity) play in the efficiency of X-ray production.
Especially, the effective temperature might be crucial as it determines the ionization state
of the atmosphere. \cite{Mohanty02.1} have argued that the electrical resistivity in the
nearly neutral atmospheres of very low-mass (VLM) dwarfs may be enhanced to the point that no efficient
coupling of matter and magnetic field is possible, preventing magnetic activity.
A possible direct dependence between X-ray luminosity and $T_{\rm eff}$ has recently been
pointed out by \cite{Stelzer06.1}. 
 
Whatever is the origin of the $L_{\rm x} - L_{\rm bol}$ relation, it ensues that 
the chance for an X-ray detection is higher for {\it young} brown dwarfs than for less 
luminous evolved ones. Indeed, in recent years deep X-ray observations with {\em Chandra} and 
{\em XMM-Newton} in star forming regions
have turned up an increasing number of X-ray detections among young brown dwarfs 
\citep[e.g. ][]{Preibisch05.2, Grosso06.1}. 

Cha\,I is relatively nearby, and therefore it was among the first star forming regions
with a well-defined brown dwarf population. 
For the median age of Cha\,I members ($\sim 2$\,Myr), 
the substellar limit corresponds to spectral type M6 to M6.5 \citep{Baraffe98.1}. 
The sample of brown dwarfs and candidate brown dwarfs in Cha\,I 
identified in an H$\alpha$ emission line survey \citep{Comeron00.1}, 
yielded the first X-ray detection of a substellar object 
with {\em ROSAT} \citep{Neuhaeuser98.1}.
In a more recent {\em XMM-Newton} observation of the southern part of Cha\,I,  
that includes all $13$ H$\alpha$ emitting
brown dwarfs and brown dwarf candidates from the survey of \cite{Comeron00.1}, 
the X-ray detection fraction for the faint ``ChaH$\alpha$'' objects was increased 
from $6/13$ to $9/13$ thanks to the higher sensitivity 
and higher spatial resolution of {\em XMM-Newton} with respect to 
{\em ROSAT} \citep{Stelzer04.2}.
Further X-ray studies with {\em Chandra} and {\em XMM-Newton}
concentrating on the northern part of the Cha\,I cloud
have been presented by \cite{Feigelson04.2} and \cite{Robrade07.1}. 

The X-ray detected and non-detected objects from these surveys 
are distinguished by different plotting symbols in Fig.~\ref{fig:hrd} 
(see figure caption).
They have been placed in the HR diagram 
using the effective temperatures and luminosities given by \cite{Luhman04.3}. 
With respect to previous estimates for the physical parameters, 
some of the objects that used to be considered brown dwarfs 
have now moved across the substellar boundary into the stellar domain;
see Sect.~\ref{sect:discussion} for more details. 
It is evident from Fig.~\ref{fig:hrd}
that the X-ray detection rate beyond the substellar boundary is rather low.
According to Luhman's compilation of physical parameters and the \cite{Chabrier00.2} models,
only three bona-fide brown dwarfs and two objects at the border line were
detected in X-rays: ChaH$\alpha$\,1, ChaH$\alpha$\,7, CHSM-17173,
ISO-217, and ESOH$\alpha$566.  
On the other hand, all Cha\,I members with $M > 0.2\,M_\odot$ that were in the field-of-view of
one of the above-mentioned X-ray observations were detected. 

All three X-ray detected bona-fide brown dwarfs in Cha\,I seem to be very young, 
located above the $1$\,Myr isochrone.
The two components of the binary 2M1101\,AB have similarly young age
and bracket 
ChaH$\alpha$\,1, ChaH$\alpha$\,7, and CHSM-17173 in terms of mass and
effective temperature.
Here we discuss recent {\em Chandra} and {\em XMM-Newton} observations of 2M\,1101. 
The observations and data analysis are described in Sect.~\ref{sect:observations}, 
and Table~\ref{tab:obslog} provides the observing log.
While {\em XMM-Newton} provides higher collecting area, only {\em Chandra} has the potential
to separate the binary and to compare the X-ray properties of 
the two components of this benchmark substellar twin. 
The X-ray properties of 2M\,1101\,A and~B are discussed in Sect.~\ref{sect:results}. 
In Sect.~\ref{sect:discussion} the results are combined 
with the published X-ray data for other low-mass Cha\,I members. 
This allows us to investigate the dependence of $L_{\rm x}$ on physical parameters such
as bolometric luminosity, effective temperature and mass for the brown dwarf population
in Cha\,I, and to compare it to similar studies in the Orion and Taurus star forming
regions. 
%
%
\begin{table}
\begin{center}
\caption{Observing log for the X-ray observations of 2M\,1101}
\label{tab:obslog}
\begin{tabular}{l|rr}\hline
Mission/Instrum.     & {\em XMM}/EPIC-pn   & {\em Chandra}/ACIS-S \\ 
\hline
Target               & Cha\,I Field\,G     & 2M\,1101-7732 \\
PI                   & M.G\"udel           & B.Stelzer \\
Obs-ID               & 0152460301          & 6396 \\
Date [UT]            & 2002-04-09@09:39:20 & 2006-05-15@04:36:46 \\
Exposure [ksec]      & 34                  & 40 \\
\hline
\end{tabular}
\end{center}
\end{table}

\section{Observations and Data Analysis}\label{sect:observations}

\subsection{Chandra}\label{subsect:chandra}

2M\,1101 was observed for $40$\,ksec with 
{\em Chandra}'s Advanced CCD Imaging Spectrometer for Spectroscopy 
(ACIS-S); see \citet{Weisskopf02.1} for details on the satellite and its instruments. 
The data analysis was carried out using the CIAO software 
package\footnote{CIAO is made available by the CXC and can be downloaded 
from \\ http://cxc.harvard.edu/ciao/download-ciao-reg.html} version 3.3.0.1. 
We started our analysis with the level\,1 events file provided by the
{\em Chandra} X-ray Center (CXC). 
In the process of converting the level\,1 events file to a level\,2 events file
we performed the following steps.  
A correction for the charge transfer inefficiency has been applied. 
We removed the pixel randomization which is automatically applied by the CXC pipeline
in order to optimize the spatial resolution. 
We filtered the events file for event grades
(retaining the standard {\em ASCA} grades $0$, $2$, $3$, $4$, and $6$), 
and applied the standard good time interval (GTI) file. 
Events flagged as cosmic rays
were retained after inspection of the images revealed that a substantial
number of source photons erroneously carry this flag. 
Since the positional accuracy is particularly
important to our observation we also checked the astrometry for any known 
systematic aspect offset using the CIAO aspect 
calculator\footnote{see http://asc.harvard.edu/ciao/threads/arcsec\_correction }. 
This tool confirms that no offset larger than $1^{\prime\prime}$ is present. 

Source detection was performed with the {\sc wavdetect} algorithm \citep{Freeman02.1} 
on an image with $0.25^{\prime\prime}$ pixel size 
centered on the position of the primary 2M\,1101\,A
and excluding photons with energy outside the $0.5-8$\,keV band.    
{\sc wavdetect} correlates the data with a mexican hat function
to search for deviations from the background. This method 
is well suited for separating closely spaced point sources.  
We used wavelet scales between $1$ and $8$ in steps of $\sqrt{2}$. 

\begin{figure}
\begin{center}
\resizebox{9cm}{!}{\includegraphics{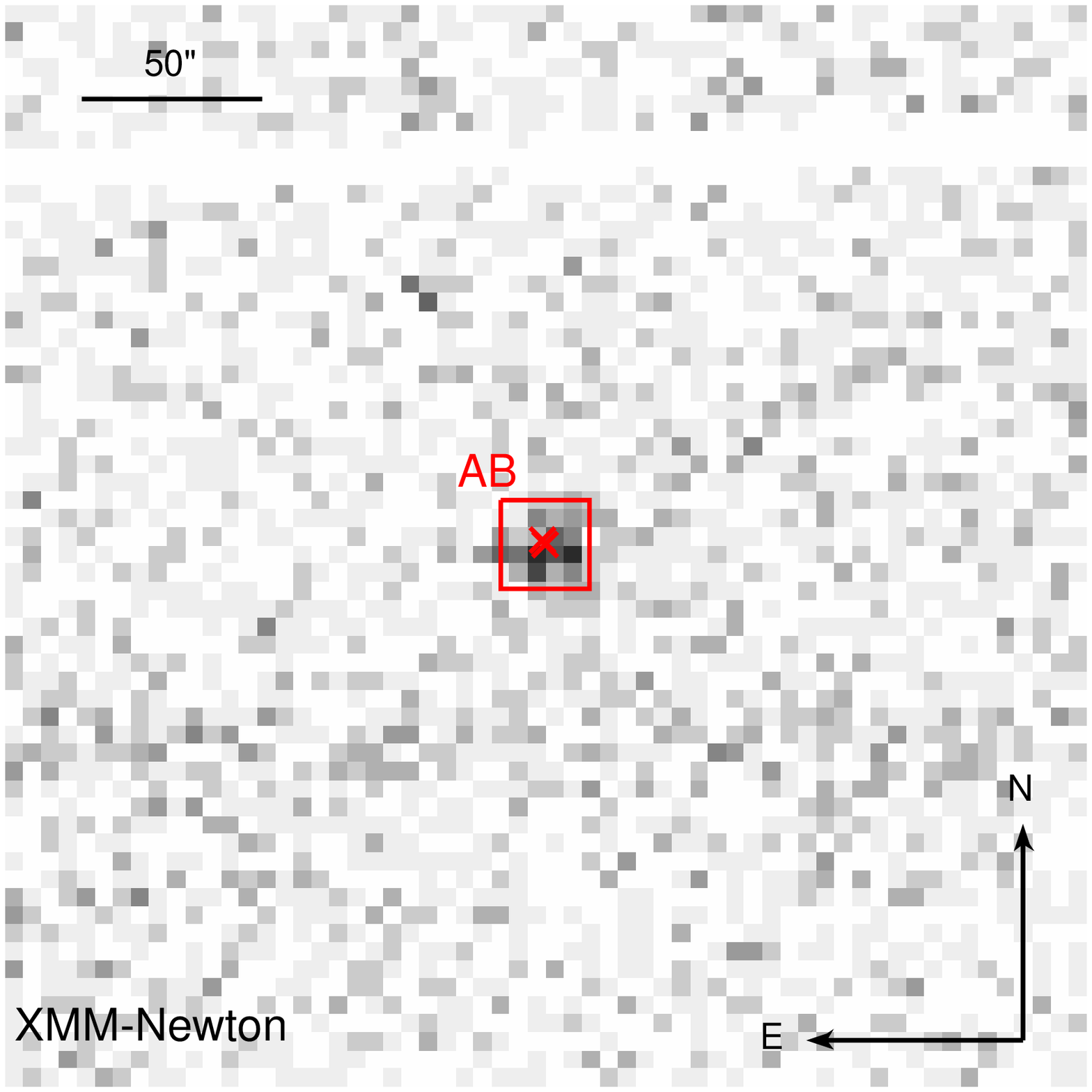}}

\vspace*{-0.6cm}
\resizebox{9cm}{!}{\includegraphics{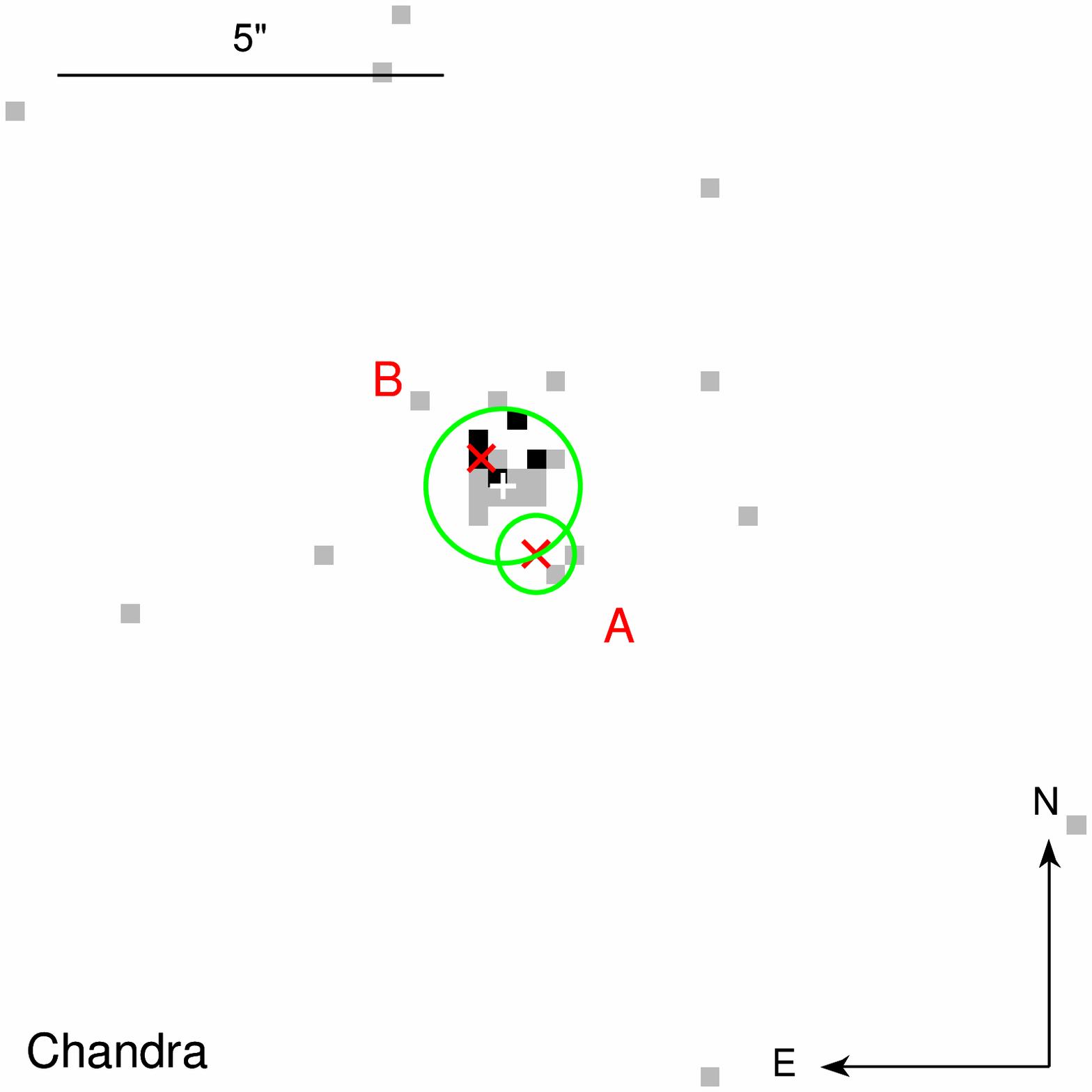}}
\caption{X-ray images of 2M\,1101 in the $0.5-8$\,keV band: 
{\em XMM-Newton} EPIC/pn in the top panel with $5^\prime \times 5^\prime$ image size, 
$5^{\prime\prime}$ pixel size and offaxis angle of $6.6^{\prime}$ for 2M\,1101. 
The white horizontal line near the top of the EPIC/pn image is a CCD gap. 
The grey square indicates the position and size
of the {\em Chandra} ACIS-S image shown in the bottom panel.   
Image size for {\em Chandra} is $24.6^{\prime\prime} \times 24.6^{\prime\prime}$,
bin size is $0.25^{\prime\prime}$, and 2M\,1101 is on-axis.   
X-points in the {\em XMM-Newton} and in the {\em Chandra} image 
denote the near-IR positions given by \protect\cite{Luhman04.3} for 
the individual components in the binary system. The circles in the
{\em Chandra} image represent the extraction areas; see text in 
Sect.~\ref{subsect:chandra}.
} 
\label{fig:images}
\end{center}
\end{figure}

We detect one X-ray source at the position of the binary. Despite an obvious
offset with respect to the near-IR position (Fig.~\ref{fig:images}), we are confident that 
this source can be identified with component~B. 
This assertion is supported by a check of the astrometry using the only other nearby
X-ray star in the field; for details see Sect.~\ref{sect:results}. 
There is no detected source associated with 2M\,1101\,A, but a possible enhancement
of the count rate may evidence weak X-ray emission from this component. 

For further analysis we extracted photons assigned to 2M\,1101\,B 
from a circular region of $1.0^{\prime\prime}$ radius centered on the 
X-ray position. That area includes $91$\,\% of the PSF. 
Photons assigned to 2M\,1101\,A are extracted from a circular region 
of $0.5^{\prime\prime}$ radius centered on its near-IR position. This latter region
includes $66$\,\% of the PSF. We note, that the extraction areas of components A and B
overlap, but there are no photons in the overlapping region; cf. Fig.~\ref{fig:images}. 
The background is very low (cf. Sect.~\ref{sect:results}) and can be neglected. 
%
%

\subsection{XMM-Newton}\label{subsect:xmm}

2M\,1101 is in the field-of-view of a 30\,ksec {\em XMM-Newton}\footnote{The satellite and its instruments
are described in a Special Issue of A\&A (2001, vol.365).} observation. 
We extracted the data taken with the 
European Photon Imaging Camera pn CCD (EPIC/pn) from the {\em XMM-Newton} archive. 
The observations were performed in full-frame mode employing the medium filter. 

Starting from the pipeline products, we filtered the events file for 
pixel patterns (retaining only singles and doubles), 
edge effects at the boundary of individual CCD chips, events outside the
field-of-view, and near bad pixels. We also eliminated the lowest pulse height channels 
to further reduce the noise. 
We searched for times of high background by selecting the uncontaminated time intervals 
with a filter that maximizes the 
signal-to-noise as a function of the count rate across the whole 
detector array and the exposure time. 
The data turned out to be widely unaffected by background flaring, with the exception 
of a brief time interval near the middle of the observation that was removed from 
the GTIs. This way the effective observing time is reduced to $28.6$\,ksec.  
In the subsequent analysis only photons that arrived within the GTIs were considered. 

2M\,1101\,AB is clearly detected, 
but as the spatial resolution
of EPIC is well above the binary separation, the two components can not
be resolved (see Fig.~\ref{fig:images}). 
We extracted source photons from a circle with $30^{\prime\prime}$ radius 
around the position of 2M\,1101\,A. According to the {\em XMM-Newton} User's 
Handbook\footnote{see http://xmm.vilspa.esa.es/external/xmm$\_$user$\_$support/documen-
tation/uhb/XMM$\_$UHB.html}  
this radius comprises $\sim 80$\,\% encircled energy. 
Background photons were extracted from two $30^{\prime\prime}$
circles positioned on the same CCD chip, at the same distance from the read-out node
as the source extraction area and avoiding the read-out strip of a nearby bright X-ray source. 
For the spectral analysis we generated a response matrix and ancilliary response file 
using standard {\em XMM-Newton} Science Analysis System tools. 
The spectrum was binned to a minimum of $15$ counts per bin
and analysed in the XSPEC v.11.3.0 environment.

\section{Results}\label{sect:results}

\begin{table*}[t]
\begin{center}
\caption{X-ray properties of 2M\,1101\,A and~B. For {\em XMM-Newton} we list results for the time-average, flare (f) and quiescent (q) state.}
\label{tab:xray}
\begin{tabular}{lrrrrrrlrr}
\hline
Component & $\Delta_{\rm ix}^{(1)}$ & Net Counts$^{(2,3)}$ & Net Rate$^{(2,3)}$             & $N_{\rm H}$                 & $kT$        & $\chi^2_{\rm red}$ & (d.o.f.) & $f_{\rm x}^{(2,4)}$   & $\log{L_{\rm x}}^{(2,4)}$ \\
          & [$^{\prime\prime}$]     &                      & \multicolumn{1}{c}{[cts/s]}  & [${\rm 10^{21}\,cm^{-2}}$]  & [keV]       &                    &          & [${\rm erg/cm^2/s}$]     & [erg/s]                   \\
\hline
\multicolumn{10}{c}{{\em XMM-Newton} EPIC/pn}\\
\hline
A+B       &        & $148 \pm 13.2$  & $(5.7 \pm 0.5)\cdot10^{-3}$ & $2.8^{4.4}_{1.7}$ & $0.9^{1.4}_{0.6}$ & $0.9$ & $ (13) $ & $3.0 \cdot 10^{-14}$ & 29.0 \\
 (f)      &        & $73.5 \pm 9.6$ & $(1.0 \pm 0.1)\cdot10^{-2}$ & $3.8^{6.1}_{0.7}$ & $0.9^{2.0}_{0.5}$ & $1.6$ & $ (6)  $ & $6.2 \cdot 10^{-14}$ & 29.3 \\
 (q)      &        & $74.5 \pm 9.7$ & $(4.0 \pm 0.5)\cdot10^{-3}$ & $3.1^{6.4}_{1.7}$ & $0.6^{1.1}_{0.4}$ & $1.0$ & $ (12) $ & $2.2 \cdot 10^{-14}$ & 28.8 \\
\hline
\multicolumn{10}{c}{{\em Chandra} ACIS-S$^{(5)}$}\\
\hline
A         &  $1.1$ & $ 2 \pm 2.7$ & $(5.1 \pm 6.8)\cdot 10^{-5}$ & $=2.5$  & $=0.9$ & $$ & $$ & $< 1.4\cdot 10^{-15}$ & $< 27.6$ \\
B         &  $0.5$ & $21 \pm 5.6$ & $(5.3 \pm 1.4)\cdot 10^{-4}$ & $=0.0$  & $=0.9$ & $$ & $$ & $  2.1\cdot 10^{-15}$ & $  27.8$ \\
\hline
\multicolumn{10}{l}{$^{(1)}$ Error of the offset derived from the uncertainty of the X-ray position is $\approx 0.1^{\prime\prime}$ for ACIS;}\\
\multicolumn{10}{l}{$^{(2)}$ in [0.5,8.0]\,keV;}\\
\multicolumn{10}{l}{$^{(3)}$ errors are computed using the Gehrels description ($1 + \sqrt{N+0.75}$).} \\
\multicolumn{10}{l}{$^{(4)}$ fluxes and luminosities are corrected for encircled PSF fraction and for absorption; the upper limit is based on the $95$\,\% confidence value} \\
\multicolumn{10}{l}{derived as described in the text.}\\ 
\multicolumn{10}{l}{$^{(5)}$ Numbers preceeded by `=' are assumed values, see text.}\\
\end{tabular}
\end{center}
\end{table*}

\begin{figure}
\begin{center}
\resizebox{9cm}{!}{\includegraphics{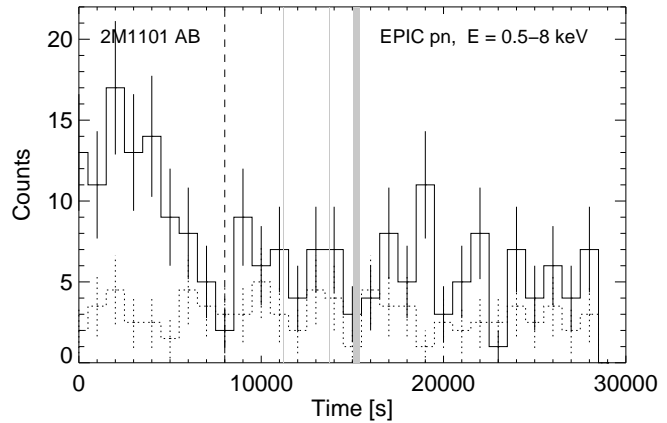}}
\caption{{\em XMM-Newton} EPIC/pn lightcurve of 2M\,1101\,AB: {\em solid line} - Signal from the position
of 2M\,1101\,AB comprising source and background, {\em dotted line} - Background from source-free
regions as described in the text. The grey shades indicate `bad time intervals' removed from 
the data, and the dashed line marks the point chosen to distinguish 
flaring and quiescent state.}
\label{fig:epicpn_lc}
\end{center}
\end{figure}

We begin the discussion of the results with the {\em XMM-Newton} data, where
the statistics are higher. The EPIC/pn lightcurve extracted from the position 
of 2M\,1101\,AB is shown in Fig.~\ref{fig:epicpn_lc} together with the area-scaled 
background lightcurve. A phase of enhanced source signal at the beginning of the observation 
is distinguished. 
The background is constant throughout the observation, and can not be responsible for
this flare-like feature. A Kolmogorov-Smirnov (KS) test for the
source+background lightcurve yielded a probability for variability $P>99$\,\%,
while no significant variability is detected with the KS test in the background-only lightcurve. 
The mean net source count rate of the combined emission from 2M\,1101\,AB measured with EPIC/pn
is given in Table~\ref{tab:xray}. 
We further split the observation in a flaring and a quiescent part; the dividing line is
indicated in Fig.~\ref{fig:epicpn_lc}. Net source count rates for the two activity states
are also listed in Table~\ref{tab:xray}.

\begin{figure}
\begin{center}
\resizebox{9cm}{!}{\includegraphics{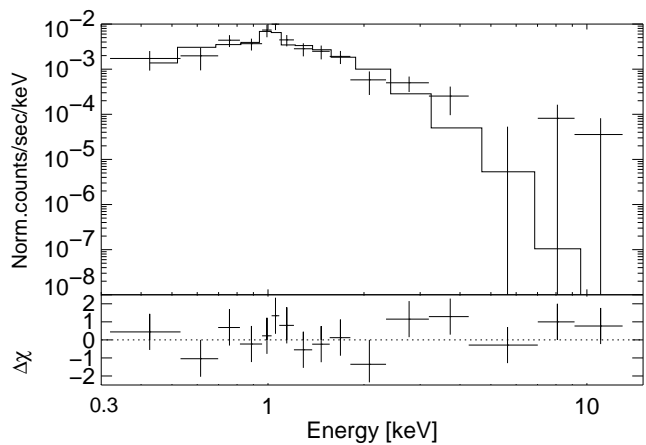}}
\caption{Time-averaged {\em XMM-Newton} EPIC/pn spectrum of the unresolved binary system 2M\,1101\,AB: data, bestfit model and residuals.}
\label{fig:epicpn_spec}
\end{center}
\end{figure}

The time-averaged 
EPIC/pn spectrum is displayed in Fig.~\ref{fig:epicpn_spec}. Albeit stellar
coronae are known to be composed of a multi-temperature plasma, in the case of poor statistics the 
X-ray spectra are often approximated fairly well by a one-temperature thermal model describing
a plasma in collisional ionization equilibrium with a  
photoabsorption term \citep[e.g.][]{Getman05.1}. 
We adopt the {\sc apec} code \citep{Smith01.1} with the {\sc wabs} absorption as
implemented in XSPEC (v.11.3). 
For the elemental abundances of the X-ray emitting plasma 
we use the values presented by \cite{Maggio07.1}. 
This set of abundances was determined from fitting the X-ray spectra of 
a sample of X-ray bright pre-MS stars in Orion. 
The major deviation with respect to photospheric abundances \citep[e.g.][]{Anders89.1, Asplund05.1} 
is the low iron abundance and the high Ne/Fe ratio, a pattern that has repeatedly been found
in stellar coronae \citep[see e.g. review by ][]{Guedel04.2}.  
The bestfit of such a model to the observed time-averaged EPIC/pn spectrum of 2M\,1101\,AB 
has $\chi^2_{\rm red} = 0.9$ with $13$ degrees of freedom. 
The bestfit parameters for average, flaring and quiescent spectrum 
are listed in Table~\ref{tab:xray}.  
During the phase of high activity at the beginning of the observation 
the temperature is higher than during the following quiescent phase
(although not statistically significant), as is typical for stellar flares. 

The spectral fit of the time-averaged EPIC/pn spectrum yields 
an absorbing column $N_{\rm H}$ (cf. Table~\ref{tab:xray}) that corresponds to a 
%
%
near-IR extinction $A_{\rm J} \approx 0.29...0.84$\,mag using the gas-to-dust conversion 
law derived by \cite{Vuong03.1} for pre-MS stars in $\rho$\,Oph. This range of $A_{\rm J}$ 
includes both the observational $90$\,\% error of $N_{\rm H}$ and the uncertainty in the
$N_{\rm H}/A_{\rm J}$ relation.
The near-IR extinction estimated this way is consistent with the {\em observed} $A_{\rm J}$ 
of 2M\,1101\,A but higher than the value of 2M\,1101\,B (cf. Table~\ref{tab:optir}).
This result suggests that the emission is dominated by component~A. 
%
%

Obviously, there is no way to separate the contributions of 2M\,1101\,A and~B to the
X-ray spectrum and lightcurve observed with {\em XMM-Newton}, and we resort to the {\em Chandra}
data where the two components can be resolved. 
Contrary to the preliminary conclusion from the {\em XMM-Newton} spectrum, 
the secondary, 2M\,1101\,B, is much closer to the 
X-ray source detected with {\sc wavdetect} than the primary, 2M\,1101\,A.  
Fig.~\ref{fig:images} suggests a shift of the X-ray image with respect to the near-IR positions 
in the south-west direction. Other than centering the X-ray source onto the position
of 2M\,1101\,B, such a shift would clearly identify 2M\,1101\,A with the nearby `cluster' of 
two photons. 
The image can not be checked for a systematic boresight error because 
there is not a sufficient number of X-ray bright sources with optical/IR counterparts. 
However, another X-ray emitting Cha\,I member (CS\,Cha) is nearby ($3.7^\prime$ off-axis). 
We compared the position of the corresponding X-ray source to the
optical position of CS\,Cha, and found an offset of only $0.06^{\prime\prime}$. 
Therefore, we are confident that no significant shifts are present in the X-ray image. 

\begin{figure}
\begin{center}
\resizebox{9cm}{!}{\includegraphics{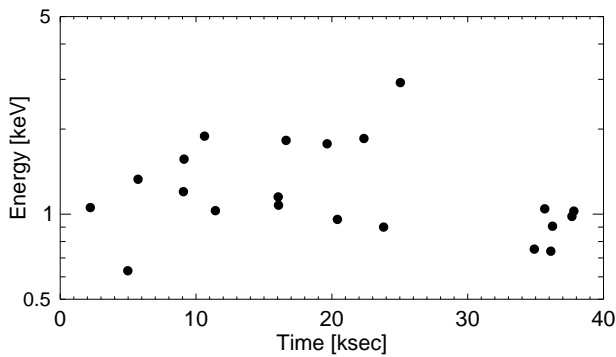}}
\caption{Photon energy vs. arrival time for the {\em Chandra} X-ray source associated with 2M\,1101\,B. 
There are no bad-time intervals.}
\label{fig:acis_ene_time}
\end{center}
\end{figure}

With a total of $21$ net counts in the extraction area 
the {\em Chandra} source associated with 2M\,1101\,B 
is too weak for spectral analysis. However, 
some information can be gained from Fig.~\ref{fig:acis_ene_time}, where we plot the energy
versus the arrival time of the individual photons in the extraction circle of 2M\,1101\,B.
%
%
The source has a remarkably hard spectrum, with
more than half of the photons having energies in excess of $1$\,keV. 
The median photon energy of 2M\,1101\,B in the $0.5-8$\,keV band is $1.06 \pm 0.17$\,keV, 
where the uncertainty was estimated with the Maritz-Jarrett method as described by \cite{Hong04.1}. 
From nearby source free regions we estimate that 
the expected number of background photons in the extraction area of 2\,M1101\,B is 
only $0.22$ in the $0.5-8$\,keV interval. 
Therefore, the background can be neglected in the evaluation of the flux in this energy band. 

Clearly, the detection of $2$ photons with energies between $0.5-8$\,keV from 2M\,1101\,A is 
significantly higher than the number of contaminating photons from the
X-ray source associated with component B. The source extraction area of 2M\,1101\,A 
comprises $3$\,\% of all photons in the PSF of 2M\,1101\,B, i.e. $0.66$\,photons.
There are $0.05$\,background counts expected for the same energy band 
in the extraction area of 2M\,1101\,A. 
Considering further the fact that the two photons are coincident with the position
of 2M\,1101\,A, they probably both originate from this brown dwarf. 
Nevertheless, local fluctuations
of the background can not be excluded, and we prefer to assign an upper limit to the count
rate of this component. 
The ``observed'' number of photons from 2M\,1101\,A 
extrapolated across the full PSF is $3$ counts.
Using the prescription by \cite{Kraft91.1} for Poisson-distributed
counting data we derive a $95$\,\% confidence upper limit of $\sim8$\,counts for 2M\,1101\,A. 

The source fluxes of 2M\,1101\,A and~B given in Table~\ref{tab:xray} were evaluated with 
PIMMS\footnote{The Portable Interactive Multi-Mission Simulator (PIMMS) is accessible at
http://asc.harvard.edu/toolkit/pimms.jsp} 
on basis of their individual near-IR extinctions and the temperature of $0.9$\,keV from 
the EPIC/pn spectrum. The sum of the fluxes of both components estimated this way from
the {\em Chandra} observation is almost a factor $10$ below the time-averaged {\em XMM-Newton} flux,
and a factor of $> 5$ below the quiescent {\em XMM-Newton} flux.
If instead the $N_{\rm H}$ value measured from the EPIC/pn spectrum is adopted for the spectral
model in PIMMS, the combined ACIS-S flux of 2M\,1101\,A and~B increases by a factor of $2$. 
Source variability may be in part responsible for the remaining discrepancy
with respect to the {\em XMM-Newton} flux. 

Indeed, in Fig.~\ref{fig:acis_ene_time} there is a striking absence of photons 
between $\sim 25...35$\,ksec from the start of the 
observation. To quantify the suspicion of source variability we applied a test proposed
by \cite{Preibisch02.1}: First, we determine the time interval $\Delta t$ 
in which $2$ photons are expected
to be detected under the assumption of a constant signal corresponding to the time averaged
count rate. Then, the maximum number of counts $N_{\rm max}$ 
in any such interval $\Delta t$ across the observation is computed. The Poisson probability
that $N_{\rm max}$ is a statistical fluctuation is given by 
$P = 1 - \sum_{k=0}^{N_{max}-1} e^{-2} \frac{2^k}{k!}$. 
We find $P = 0.016$. Therefore, 2M\,1101\,B was probably variable during the {\em Chandra} observation.

\section{Discussion}\label{sect:discussion}

\subsection{Comparison of {\em XMM-Newton} and {\em Chandra} results}\label{subsect:xrays_2m1101}

At first sight, the {\em Chandra} detection of the secondary component and the non-detection (or extreme
weakness) of the primary in the 2M\,1101 system seems to contradict the 
$L_{\rm x} - L_{\rm bol}$ relation. 
However, one must take account of the fact that the X-ray photons are subject to absorption.
\cite{Luhman04.3} measured negligible extinction for 2M\,1101\,B but $A_{\rm J} = 0.45$\,mag
for 2M\,1101\,A. If we assume that both objects have the same X-ray flux and spectrum with temperature
of $\sim 1$\,keV, we can estimate the effect of the absorbing column onto the {\em observed}
spectrum. 
With PIMMS we found that the count rate of the unabsorbed ($N_{\rm H} = 0$) X-ray source 
2M\,1101\,B corresponds to a flux of $2.1\cdot10^{-15}\,{\rm erg/cm^2/s}$. 
If the column density was 
$N_{\rm H} = 2.5 \cdot 10^{21}\,{\rm cm^{-2}}$ 
(equivalent to the observed $A_{\rm J}$ of 2M1101\,A), 
a total of $7-8$\,counts would be expected in the extraction area of 2M\,1101\,A, 
in contrast to the $2$ counts observed. 
In order to shrink the count rate to the observed level, both a lower temperature ($\approx 0.6$\,keV)
and a higher extinction ($N_{\rm H} \approx 5\cdot10^{21}\,{\rm cm^{-2}}$) 
must be hypothesized for 2M\,1101\,A. 
This is, however, unconsistent with the {\em XMM-Newton} spectrum. 
We conclude, that 
during the {\em Chandra} observation the X-ray flux of 2M\,1101\,A was significantly lower
than that of the photospherically cooler secondary 2M\,1101\,B. 

Furthermore, the {\em XMM-Newton} spectrum suggests that 
the dominant X-ray emitter during the {\em XMM-Newton} observation was 2M\,1101\,A, while during
the {\em Chandra} observation clearly the dominant X-ray emitter was 2M\,1101\,B, and consequently
both objects must be strongly variable in X-rays. 
Using the results given in Table~\ref{tab:xray} this leads to the conclusion 
that the combined X-ray luminosity of 2M\,1101\,AB can vary by at least a factor of $10$. 
An alternative interpretation is that 
the dominant X-ray emitter was 2M\,1101\,B during both observations but its 
$N_{\rm H}$ is higher than the near-IR value. In this latter case, the {\em Chandra} 
flux can be estimated using the $N_{\rm H}$ obtained from the {\em XMM-Newton} spectrum. 
This flux is $4.4 \times 10^{-15}\,{\rm erg/cm^2/s}$, a factor of two higher than for 
negligible extinction. However, it is still a factor of seven lower than
the {\em XMM-Newton} flux. Therefore, in this scenario the X-ray emission of 2M\,1101\,B, 
again, must be variable. Furthermore, the $N_{\rm H}$ value from the {\em XMM-Newton}
spectrum corresponds to a near-IR extinction of $A_{\rm J} \approx 0.5$\,mag, and this
would shift 2M\,1101\,B upward 
in the HR diagram. As a consequence, it would no longer be coeval with 2M\,1101\,A,
an unlikely scenario. 

We caution that the above considerations rely on the gas-to-dust extinction
law presented by \cite{Vuong03.1}.
Deviations from the standard extinction law 
have been seen in other samples of young stars \citep[cf.][]{Feigelson05.1}.
Such discrepancies can arise from errors in the X-ray spectral model or in the photometry,
or they may represent untypical environmental conditions.

\subsection{X-ray emission from brown dwarfs in Cha\,I}\label{subsect:xrays_chaI}

To put 2M\,1101\,A and~B into the context of the 
X-ray properties of low-mass stars, we revisit published X-ray data for the Cha\,I star
forming region. Studies of the X-ray population of Cha\,I with
{\em XMM-Newton} and {\em Chandra} 
have been presented by \cite{Stelzer04.2}, \cite{Feigelson04.2}, 
and \cite{Robrade07.1}. 
We resume here the results on the VLM stars and brown dwarfs based on the
membership catalog of \cite{Luhman04.1} with additional $9$ members discovered
by \cite{Comeron04.1}.  
Previous estimates for the physical parameters had placed some of the 
objects which now appear to be stars into the brown dwarf regime.
This was due to a combination of (i) slightly different spectral types assigned,
(ii) a different SpT / $T_{\rm eff}$ conversion and
(iii) the use of other evolutionary models.
\cite{Luhman04.1} has shown that for the Cha\,I population
the models by \cite{Baraffe98.1} and \cite{Chabrier00.2} provide the best
agreement with observational constraints, including higher-mass T\,Tauri stars 
up to $M = 1\,M_\odot$. We use these models 
and the physical parameters from \cite{Luhman04.1} and \cite{Comeron04.1},  
to estimate individual masses for all X-ray detected Cha\,I members
from the above-mentioned publications. 

Then we re-inspected these X-ray source lists for consistency with the updated membership list. 
Upper limits for non-detections were not discussed in the X-ray surveys, 
and we will not consider them here. The assessment of upper limits
is difficult because of their dependence on the spectral shape and, more importantly, 
the (unknown) extinction. We verified that, for the observations discussed here, these 
upper limits range roughly between $\log{L_{\rm x}}\,{\rm [erg/s]} < 28.5$ and $< 27.5$. 
All X-ray detected Cha\,I members are included in the X-ray catalogs. 
For the faint objects from \cite{Robrade07.1}, we computed $L_{\rm x}$ from the
listed count rates and conversion factor, assuming a distance of $160$\,pc. 
The examined fields include $75$ detected and $20$ undetected Cha\,I members. 
All non-detections have masses below $0.2\,M_\odot$,
i.e. the X-ray census in Cha\,I is complete down to this limit,
at least in the area surveyed so far which comprises the central star forming
sites. 

\begin{figure}[b]
\begin{center}
\resizebox{9cm}{!}{\includegraphics{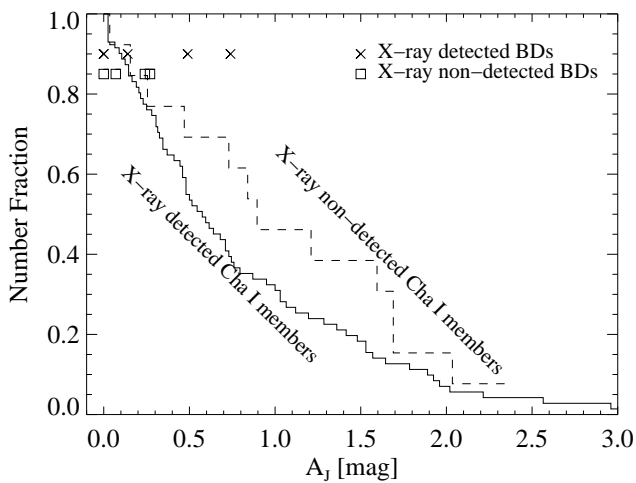}}
\caption{Comparison of $J$ band extinction for X-ray detected (solid line) and 
X-ray undetected (dashed line) Cha\,I members; $A_{\rm J}$ from \cite{Luhman04.1}. 
Substellar objects are plotted 
at their individual $A_{\rm J}$ values and at an arbitrarily chosen fixed 
value on the vertical axis:
Crosses are detected brown dwarfs, and squares undetected ones; substellar regime 
assumed to correspond to spectral type $\geq$ M6.}
\label{fig:aj}
\end{center}
\end{figure}

For the Orion Nebula Cluster (ONC) it was shown that, next to intrinsically low X-ray flux, 
high extinction is one of the causes why even deep exposures may fail to 
register X-rays from (sub)stellar objects \citep{Preibisch05.1}. 
The cumulative number distribution of the $J$ band extinction, $A_{\rm J}$, is
shown in Fig.~\ref{fig:aj} for all X-ray observed Cha\,I members. 
The $A_{\rm J}-$distribution of X-ray detections and non-detections is undistinguishable,
as we verified in $2$-sample tests implemented in ASURV\footnote{The Astrophysical SURVival analysis
package is available at the astrostatistics site at Penn State University; 
see also \protect\cite{LaValley92.1}.}.
In the sub-sample of probable brown dwarfs (spectral type $\geq$M6) 
there is also no significant distinction between   
X-ray detected and undetected ones in terms of $A_{\rm J}$, such that the non-detection
of $4$ brown dwarfs is probably due to their intrinsically weak emission.  
%
%

As mentioned in Sect.~\ref{sect:intro}, prior to the study presented here,
only three bona-fide brown dwarfs of Cha\,I had been detected in X-rays: 
ChaH$\alpha$\,1 and~7 \citep{Stelzer04.2} 
and CHSM-17173 \citep{Robrade07.1}. 
We add here the X-ray detection of 2M\,1101\,B and possibly also 2M\,1101\,A. 
With its spectral type of M8.25, 2M\,1101\,B is the coolest and least massive 
brown dwarf of Cha\,I detected in X-rays.
Its X-ray luminosity is at the low end of all X-ray detected VLM objects in Cha\,I,
and the upper limit for 2M\,1101\,A is lower than any of the detections. 

\begin{figure}[b]
\begin{center}
\resizebox{9cm}{!}{\includegraphics{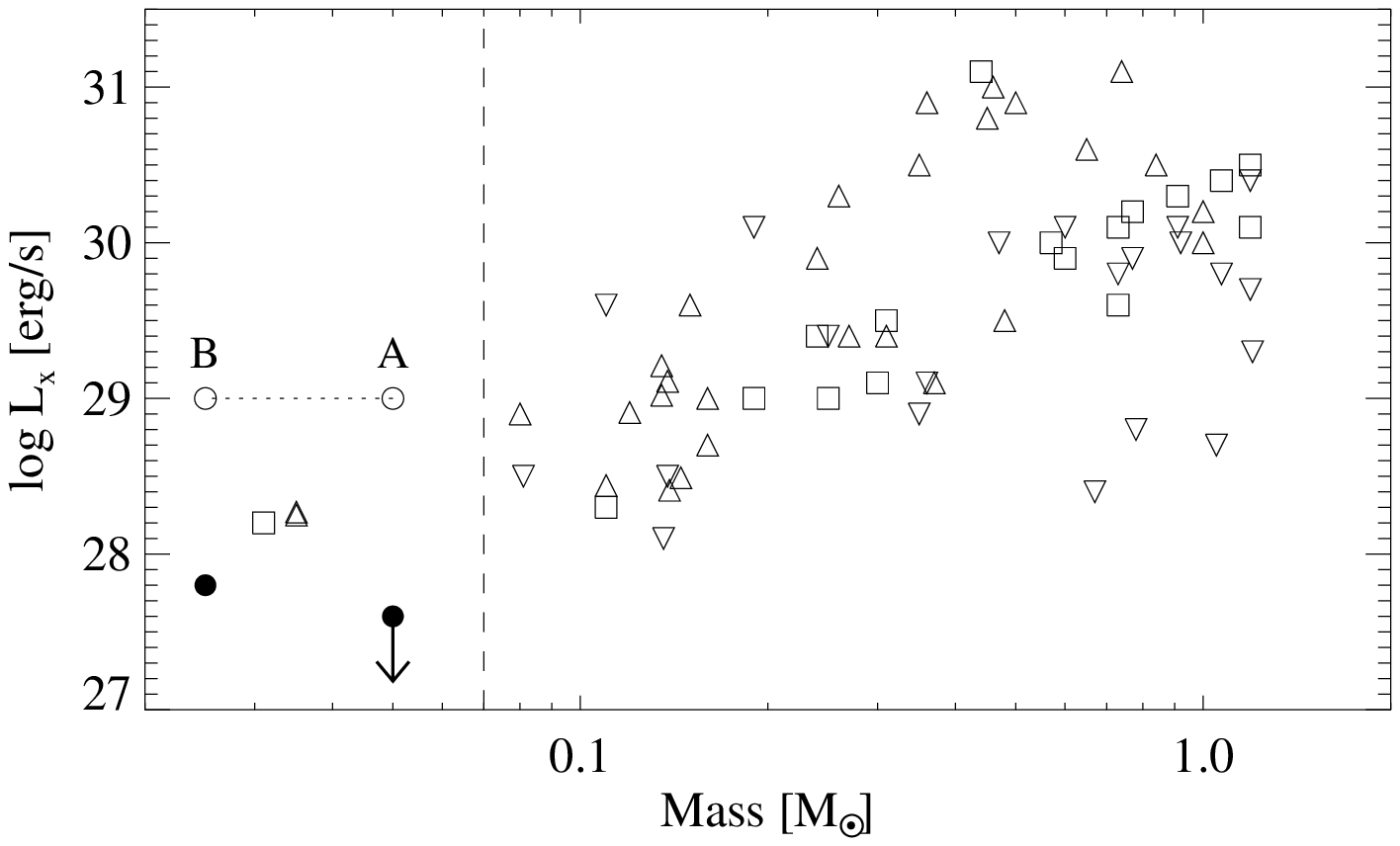}}
\resizebox{9cm}{!}{\includegraphics{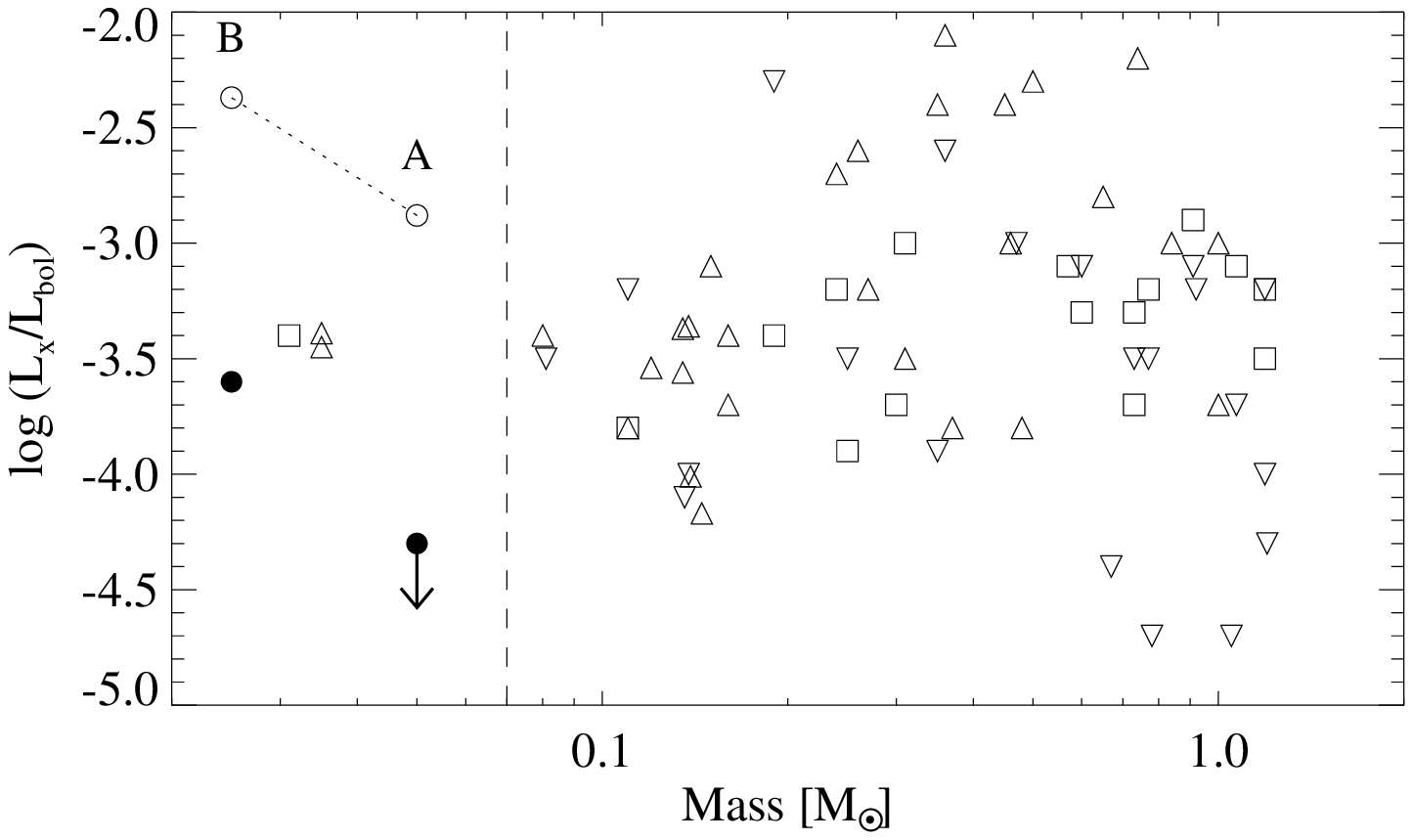}}
\caption{X-ray luminosity versus mass 
for the low-mass Cha\,I members from the catalog of 
\protect\cite{Luhman04.1}. For clarity no upper limits are shown. 
Upward triangles, downward triangles, and squares represent data from 
\cite{Stelzer04.2}, \cite{Feigelson04.2}, and \cite{Robrade07.1}, respectively. 
For 2M\,1101\,A and~B we show two positions:
(a) according to the {\em XMM-Newton} observation of the combined system 2M\,1101\,AB
(open circles connected by dashed line to symbolize that the data is from one X-ray source),
and (b) according to the {\em Chandra} observation in which the two components 
are resolved (filled circles).
The dashed lines represent the substellar limit.}
\label{fig:lx_mass}
\end{center}
\end{figure}

Fig.~\ref{fig:lx_mass} shows the dependence of X-ray emission of the young members in Cha\,I on mass.
The most important features in this diagram are 
completely consistent with findings in other star forming regions,
in particular the ONC and Taurus, and can be summarized like this: 
(i) The well-known decline of $L_{\rm x}$ with decreasing mass is seen across the low-mass range 
\citep[cf. ][for the ONC and for Taurus, respectively]{Preibisch05.1, Telleschi07.1}, 
and this trend seems to continue into the brown dwarf regime 
\citep[cf.][]{Preibisch05.2, Grosso06.1};
(ii) there is no dependence of $L_{\rm x}/L_{\rm bol}$ on mass; 
(iii) for given mass a considerable ($2-3$\,dex) spread in X-ray luminosity 
and $L_{\rm x}/L_{\rm bol}$ is observed. 
We recall, that the values shown in Fig.~\ref{fig:lx_mass} are a collection
of data obtained with different instruments and different analysis techniques. Nevertheless,
there seem not to be very significant systematic shifts of the X-ray luminosities measured 
in the three different studies.
In particular, the fields analysed by \cite{Feigelson04.2} and \cite{Robrade07.1} partially
overlap. Most stars common to both fields show typical differences of a factor of two,
which can be explained by the different energy bands in which the X-ray luminosity was measured:
For a $1$\,keV spectrum with $\log{N_{\rm H}}\,{\rm [cm^{-2}]} \approx 21.5$, 
the $0.5-8$\,keV band used by \cite{Feigelson04.2} 
includes roughly half the flux of the $0.2-10$\,keV band used by \cite{Robrade07.1}. 
The X-ray fainter objects remain undetected in the {\em XMM-Newton} pointing, because at 
a similar effective exposure time of $\sim 70$\,ksec 
{\em Chandra} is more sensitive due to its low background. 
An extensive investigation of the X-ray census in Cha\,I is beyond the scope of this
paper. Such a study is underway (Telleschi et al., in prep.), and we do not elaborate on 
this issue here. 
%
%

\begin{figure}
\begin{center}
\resizebox{9cm}{!}{\includegraphics{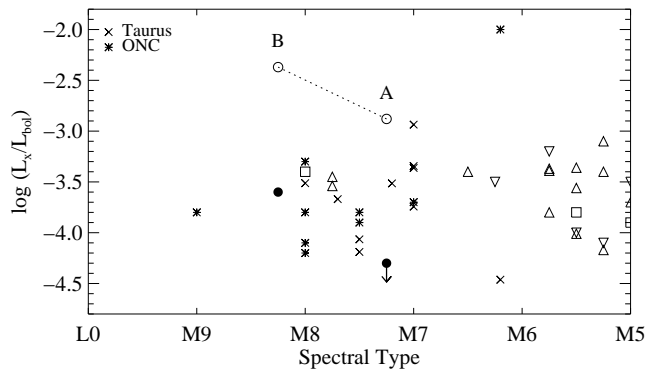}}
\caption{Fractional X-ray luminosity versus spectral type for brown dwarfs of different 
star forming regions: Cha\,I - plotting symbols as in Fig.~\ref{fig:lx_mass}, 
Taurus - crosses \citep[data from][]{Grosso06.1},
and ONC - asterisks \citep[data from][]{Preibisch05.2}.}
\label{fig:lxlbol_spt_bds}
\end{center}
\end{figure}

In Fig.~\ref{fig:lxlbol_spt_bds} the fractional X-ray luminosities of the VLM stars and
brown dwarfs in Cha\,I are complemented with equivalent objects in Taurus and the ONC. 
Similar plots have been presented individually for the ONC \citep{Preibisch05.1} and for Taurus
\citep{Grosso06.1}. 
The samples in these three star forming regions 
show that young substellar objects can emit as much X-rays as higher-mass T~Tauri stars
with respect to their bolometric luminosity, with the most active objects emitting near 
the saturation limit of $\log{(L_{\rm x}/L_{\rm bol})} \sim -3$. 
In other words, the efficiency of the dynamo and the manifestation of activity phenomena show no
break down to spectral type $\approx$M8 for objects of a few Myrs age. This is in agreement with
recent mean field dynamo models by \cite{Chabrier06.1} that predict large-scale fields in fully
convective objects, no matter whether they are stars or brown dwarfs, 
despite the absence of differential rotation in these objects. 
The X-ray observations of young brown dwarfs suggest, 
that the supposed decline of X-ray emission with decreasing atmospheric temperature 
due to the decoupling of plasma and field 
sets in at spectral types later than M8 for young ages, 
i.e. for $T_{\rm eff} < \sim 2700$\,K, consistent with the temperature range where 
the electrical resistivity is predicted to decline rapidly \cite{Mohanty02.1}. 
Previous evidence for a connection between
the X-ray emission level from brown dwarfs and $T_{\rm eff}$ regarded more evolved 
ultracool dwarfs that have effective temperatures below $2500$\,K \citep{Stelzer06.1}, and they present 
no contradiction to the picture drawn above for {\em young} substellar objects. 

Despite the obvious ability to maintain high fractional X-ray luminosities, 
the X-ray detections of young brown dwarfs reported so far represent only the tip of the 
iceberg. The bona-fide brown dwarfs in Cha\,I from published {\em Chandra} and/or {\em XMM-Newton}
observations are summarized in Table~\ref{tab:bds_xrays}. 
All Cha\,I brown dwarfs detected in X-rays so 
far are located on or above the $1$\,Myr isochrone, and a similar tendency is obvious in 
Taurus and the ONC, where mostly the very young, and consequently very luminous 
VLM stars and brown dwarfs have been detected in X-rays \citep{Preibisch05.2, Grosso06.1}. 
With the exception of 2M\,1101\,A, the undetected brown dwarfs have older ages on the
\cite{Chabrier00.2} tracks, i.e. lower bolometric luminosity. Since their upper limits to 
$L_{\rm x}$ are on average somewhat below the X-ray luminosity of the detections, 
they are expected to have $L_{\rm x}/L_{\rm bol}$ levels near the range of the X-ray detected
brown dwarfs. 
Therefore, deep X-ray pointings of low-luminosity young brown dwarfs will likely populate the
$L_{\rm x}/L_{\rm bol}$ relation with a scatter similar to that observed for higher-mass
pre-MS stars, while a shut off of X-ray emission may be expected for young brown dwarfs with
temperatures below the ones observed so far.
%
%

\begin{table}
\begin{center}
\caption{Brown dwarfs and candidates (spectral type $>$M6) of Cha\,I 
observed with {\em XMM-Newton} and {\em Chandra}.
Spectral type from \cite{Luhman04.3}, mass and age interpolated from \cite{Chabrier00.2} models.
According to \cite{Chabrier00.2} the substellar limit is located at $\sim 0.07\,M_\odot$,
which corresponds to spectral type $\approx$ M6.5 (see also Fig.~\ref{fig:hrd}). 
The X-ray luminosity is from the reference given in the last column.}
\label{tab:bds_xrays}
\begin{tabular}{llrrrl}
\hline
Designation     & SpT   & Mass        & $\log{\rm Age}$ & $\log{L_{\rm x}}$ & Ref$^*$ \\ 
                &       & [$M_\odot$] & [yrs]           & [erg/s]           & \\
\hline 
2M\,1101\,B     & M8.25 & 0.025 & $  6.0$ & $27.8$ & this work \\
ChaH$\alpha$1   & M7.75 & 0.035 & $< 6.0$ & $28.3$ & S04 \\
ChaH$\alpha$7   & M7.75 & 0.035 & $< 6.0$ & $28.3$ & S04 \\
CHSM\,17173     & M8    & 0.041 & $< 6.0$ & $28.5$ & F04 \\
ESOH$\alpha$566 & M6.5  & 0.08  & $< 6.0$ & $28.2$ & R07 \\
ISO-217         & M6.25 & 0.081 & $  6.5$ & $28.9$ & S04 \\
\hline
2M\,1101\,A     & M7.25 & 0.05  & $  6.0$ & $<27.6$ & this work \\
ChaH$\alpha$\,11 & M7.25 & 0.054 & 7.1 & $-$ & S04 \\
ISO-138         & M6.5  & 0.068 & 7.0 & $-$ & S04 \\
ChaH$\alpha$\,12 & M6.5  & 0.074 & 6.4 & $-$ & S04 \\
ChaH$\alpha$\,10 & M6.25 & 0.074 & 6.9 & $-$ & S04 \\
\hline
\multicolumn{6}{l}{$^*$ References for X-ray luminosity: S04 - \cite{Stelzer04.2},} \\
\multicolumn{6}{l}{F04 - \cite{Feigelson04.2}, R07 - \cite{Robrade07.1}.} \\
\end{tabular}
\end{center}
\end{table}

\begin{acknowledgements}
We acknowledge financial support from ASI/INAF contract I/023/05/0.
BS wishes to thank E. Flaccomio for stimulating discussions. 
\end{acknowledgements}

\bibliographystyle{aa} 
\bibliography{aa7564}

\end{document}